\documentclass[11pt,a4paper]{article}

\usepackage{amsmath,amssymb,wasysym,cite,epsfig,graphicx}

\begin{document}

\title{Gravitational lensing by a Horndeski black hole} 
\author{Javier Bad\'ia$^{1}$\thanks{e-mail: javobadia@gmail.com}, Ernesto F. Eiroa$^{1, 2}$\thanks{e-mail: eiroa@iafe.uba.ar}\\
{\small $^1$ Instituto de Astronom\'{\i}a y F\'{\i}sica del Espacio (IAFE, CONICET-UBA),}\\
{\small Casilla de Correo 67, Sucursal 28, 1428, Buenos Aires, Argentina}\\
{\small $^2$ Departamento de F\'{\i}sica, Facultad de Ciencias Exactas y Naturales,} \\ 
{\small Universidad de Buenos Aires, Ciudad Universitaria Pabell\'on I, 1428, Buenos Aires, Argentina}}

\maketitle

\begin{abstract}

In this article we study gravitational lensing by non-rotating and asymptotically flat black holes in Horndeski theory. By adopting the strong deflection limit, we calculate the deflection angle, from which we obtain the positions and the magnifications of the relativistic images. We compare our results with those corresponding to black holes in General Relativity. We analyze the astrophysical consequences in the case of the nearest supermassive black holes.

\end{abstract}

\section{Introduction}\label{intro}

Gravitational lensing by compact objects with a photon sphere presents some distinctive features. Besides the primary and secondary images, there exist two infinite sets of so-called relativistic images \cite{virbha1}, due to light rays passing close to the photon sphere, which suffer large deflection angles. The strong evidence of the presence of supermassive black holes at the center of most galaxies, including ours \cite{gillessen17} and the closest one \cite{broderick15}, has led to an important interest in this topic in recent years. The strong deflection limit allows for an analytical treatment of the relativistic images. This method, consisting in a logarithmic approximation of the deflection angle for light rays deflecting close to the photon sphere, was introduced for the Schwarzschild black hole \cite{darwin-otros,bozza01}, extended to the Reissner--Nordstr\"om geometry \cite{eiroto}, and generalized to arbitrary spherically symmetric and asymptotically flat spacetimes \cite{bozza02}. It has been recently improved by simplifying the form to perform the calculations \cite{tsukamoto17}. Within this limit, one can analytically obtain the positions, the magnifications, and the time delays of the relativistic images. Many articles that consider strong deflection lenses with spherical symmetry can be found in the literature \cite{nakedsing1,nakedsing2,retro,lenseq,virbha2,alternative,bwlens,scalarlens,tsukamoto-wh}, in the context of general relativity and also in modified gravity; most of them are analytical and use the strong deflection limit. The lensing effects of rotating black holes have also been analyzed \cite{rotbh} in the  last decade. It is expected that the observation of some optical effects in the vicinity of the supermassive black holes in the Milky Way and also in nearby galaxies, including direct imaging, will be possible in the near future \cite{observ}.

The unsolved problem of the nature of the dark matter and the dark energy necessary for the explanation of the observed features of the Universe within the context of general relativity has led to a growing interest in modified gravity theories. Among them, the scalar-tensor theories, in which an additional scalar degree of freedom is added, provide the simplest extension of general relativity. Four decades ago, Horndeski \cite{horndeski} found the most general scalar tensor theory with second order derivative equations of motion. In this theory, the action has the form
\begin{equation}
S=\int d^{4}x\sqrt{-g}\left( \mathcal{L}_2 + \mathcal{L}_3 + \mathcal{L}_4 + \mathcal{L}_5 \right) ,
\label{eq:action1}
\end{equation}
with
\begin{equation}
\mathcal{L}_2=G_2,  
\quad
\mathcal{L}_3=-G_{3}\square\phi, 
\quad
\mathcal{L}_4=G_{4}R+G_{4X}\left[(\square\phi)^2-(\nabla_\mu \nabla_\nu  \phi)^2\right], \nonumber
\end{equation}
\begin{equation}
\mathcal{L}_5=G_{5}G_{\mu\nu}\nabla^\mu \nabla^\nu  \phi-\frac{G_{5X}}{6}\left[(\square\phi)^3-3\square\phi(\nabla_\mu \nabla_\nu  \phi)^2 +2(\nabla_\mu \nabla_\nu  \phi)^3 \right], \nonumber
\end{equation}
where $g_{\mu\nu}$ is the metric tensor and $g\equiv {\rm det}(g_{\mu\nu})$ its
determinant, $R$ and $G_{\mu\nu}$ denote the Ricci scalar and the Einstein tensor, respectively; the functions $G_{i}=G_{i}(\phi,X)$ depend only on the scalar field $\phi$ and its kinetic energy $X=-\partial_\mu\phi\partial^\mu\phi/2$, while the subscript $X$ stands for the derivative with respect to $X$.  We have also used the abbreviated notation
$\square\phi\equiv g^{\mu\nu} \nabla_\mu \nabla_\nu  \phi$, $(\nabla_\mu \nabla_\nu  \phi)^2 \equiv \nabla_\mu \nabla_\nu  \phi\nabla^\nu \nabla^\mu  \phi$, and $(\nabla_\mu \nabla_\nu  \phi)^3 \equiv \nabla_\mu \nabla_\nu  \phi \nabla^\nu \nabla^\alpha \phi \nabla_\alpha \nabla^\mu  \phi$. Horndeski gravity includes quintessence, k-essence, and $f(R)$ theories as particular cases. Since a few years ago, there has been a revival of Horndeski theory \cite{horndeski-rev}, being the main theoretical framework for scalar-tensor models in which cosmological observations can be interpreted. Recently, a new class of scalar-tensor theories that extend Horndeski (dubbed ``beyond Horndeski") was introduced, with equations of motion of higher order in the derivatives \cite{glpv15}, but with the property that the true propagating degrees of freedom obey well-behaved second-order equations, thus being free from Ostrogradski instabilities. The study of black holes in Horndeski and beyond Horndeski theories has received much attention lately \cite{horndeski-bh,bcl16,bcl17}. Other researchers have explored various related topics in these theories \cite{horndeski-other}. 

In this work, we investigate the behavior as gravitational lenses of the spherically symmetric and asymptotically flat black holes in Horndeski gravity introduced in Ref. \cite{bcl17}. In Sec. \ref{sdl}, we review the main properties of the spacetime, we approximate the deflection angle by using the strong deflection limit, and we introduce the lens equation to find the analytical expressions for the positions and the magnifications of the images. We also calculate the corresponding observables. In Sec. \ref{astro}, we apply the formalism to the cases of the supermassive black hole at SgrA* and the supermassive black hole in M87. Finally, in Sec. \ref{sum}, we summarize and discuss the results obtained. We adopt units with $G=c=1$.

\section{Black hole lens in Horndeski gravity}\label{sdl}

We consider the particular case of the action (\ref{eq:action1}) in which $G_2=\eta X$, $G_4=\zeta +\beta \sqrt{-X}$, and $G_3=G_5=0$, where $\eta$ and $\beta$ are dimensionless parameters and $\zeta = M_{\rm Pl}^2/(16\pi)$. Then, the action takes the explicit form
\begin{equation}
S=\int d^{4}x \sqrt{-g} \left\{ \left[ \zeta + \beta \sqrt{(\partial \phi )^2/2} \right] R - \frac{\eta}{2} (\partial \phi )^2 - \frac{\beta}{\sqrt{2(\partial \phi )^2}}\left[ (\square \phi)^2 - (\nabla_\mu \nabla_\nu  \phi)^2 \right]\right\} .
\label{eq:action2}
\end{equation}
The coefficient $\zeta $ gives the Einstein-Hilbert part of the action; one of the parameters $\eta$ and $\beta$ can be absorbed into the scalar field by means of a redefinition, but we will not do it in order to trace the origin of the different terms. The field equations resulting from Eq. (\ref{eq:action2}) admit a static, spherically symmetric, and asymptotically flat solution \cite{bcl17} of the form 
\begin{equation}
ds^2 = -A(r) dt^2 + B(r) dr^2 + C(r)(d\theta^2 + \sin^2 \theta\, d\phi^2)
\label{eq:metric}
\end{equation}
with
\begin{equation}
A(r) = 1 - \frac{\mu}{r} - \frac{\beta ^2}{2 \zeta \eta r^2}, \quad B(r) = \frac{1}{A(r)}, \quad C(r) = r^2.
\label{eq:mfr}
\end{equation}
The integration constant $\mu$ can be interpreted as twice the black hole mass, i.e. $\mu = 2M$. The parameters $\beta $ and $\eta$ should share the same sign \cite{bcl17}, and the scalar field is given by
\begin{equation}
\phi (r) = \pm 2 \sqrt{\frac{\zeta}{\eta}} \left\{ \mathrm{Arctan} \left[\frac{\beta ^2 + \zeta \eta \mu r}{\beta \sqrt{2\zeta \eta r(r-\mu)-\beta ^2}}\right] - \mathrm{Arctan} \left( \frac{\mu}{\beta} \sqrt{\frac{\zeta \eta }{2}} \right) \right\}
\quad \mathrm{if} \; \beta >0 \; \mathrm{and} \; \eta >0,
\label{eq:phipos}
\end{equation}
\begin{equation}
\phi (r) = \pm 2 \sqrt{\frac{\zeta}{-\eta}} \left\{ \mathrm{Arcth} \left[\frac{\beta ^2 + \zeta \eta \mu r}{\beta \sqrt{\beta ^2 - 2\zeta \eta r(r-\mu)}}\right] + \mathrm{Arcth} \left( \frac{\mu}{\beta} \sqrt{\frac{-\zeta \eta }{2}} \right) \right\}
\quad \mathrm{if} \; \beta <0 \; \mathrm{and} \; \eta <0.
\label{eq:phineg}
\end{equation}
The geometry is singular at the origin of coordinates. It is useful to define the parameter $\gamma = \beta ^2/ (2 \zeta \eta )$ in order to simplify the notation. We can see that a negative value of $\gamma$ makes this metric identical to the Reissner-Nordstr\"om metric, with the squared charge given by $Q^2 = -\gamma$. Adimensionalizing with $\mu$, we let $x = r / \mu$, $\tilde{t} = t / \mu$, and $\tilde{\gamma} = \gamma / \mu^2$, so that the metric functions become
\begin{equation}
A(x) = 1 - \frac{1}{x} - \frac{\tilde{\gamma}}{x^2}, \quad B(x) = \frac{1}{A(x)}, \quad C(x) = x^2.
\label{eq:mfx}
\end{equation}
The radius of the event horizon is obtained as the largest solution of the equation $A(x)=0$, to give $x_h = (1+\sqrt{1+4\tilde{\gamma}})/2$. Therefore, to avoid a naked singularity we must have $\tilde{\gamma} \ge -1/4$, with $\tilde{\gamma} = -1/4$ in the case of an extremal black hole. Correspondingly, we work with $\tilde{\gamma}$ within the interval $[-1/4, +\infty)$.

\subsection{Deflection angle: strong deflection limit}\label{angle}

We analyze the lensing effects produced by the geometry (\ref{eq:metric}) in the case of photons passing very close to the photon sphere, which has a radius $x_m$ determined by the largest positive solution of the equation
\begin{equation}
\frac{C'(x)}{C(x)} = \frac{A'(x)}{A(x)},
\end{equation}
where the prime stands for the derivative. By replacing the metric functions (\ref{eq:mfx}), this equation reduces to 
\begin{equation}
2 =  \frac{x + 2\tilde{\gamma}}{x^2 - x - \tilde{\gamma}},
\end{equation}
which yields
\begin{equation}\label{eq:xm}
x_m = \frac{3 + \sqrt{9 + 32\tilde{\gamma}}}{4}.
\end{equation}
The deflection angle $\alpha$ for a light ray passing by the black hole with a closest approach distance $x_0$ is \cite{weinberg,nakedsing1}
\begin{equation}\label{eq:alphaint}
\alpha(x_0) = 2 \int_{x_0}^\infty\mathrm{d}x\ \sqrt{\frac{B(x)}{C(x)R(x)}} - \pi \equiv I(x_0) - \pi,
\end{equation}
where
\begin{equation}\label{eq:r}
R(x) = \frac{C(x) A_m}{A(x) C_m} - 1
\end{equation}
with $A_m \equiv A(x_m)$ and similarly for any other function of the radial coordinate. We follow the method described in Ref. \cite{tsukamoto17}, which is in turn an improvement to the one developed in Ref. \cite{bozza02}, in order to approximate the deflection angle $\alpha$. As $x_0$ approaches $x_m$ this angle diverges logarithmically, and it can be approximated in the form \cite{bozza02,tsukamoto17}
\begin{equation}\label{eq:alpha1}
\alpha(x_0) = - a_1 \log \left( \frac{x_0}{x_m} - 1 \right) + a_2 + O\left( (x_0-x_m)^2\log(x_0-x_m) \right).
\end{equation}
It is more convenient in what follows to express $\alpha$ in terms of the impact parameter $b$ for a given light ray. The relation between $b$ and $x_0$ is given by \cite{weinberg,nakedsing1}
\begin{equation}\label{eq:bx0}
b(x_0) = \sqrt{ \frac{C(x_0)}{A(x_0)} }
\end{equation}
which, expanded to second order in $x_0 - x_m$ and substituted into Eq. (\ref{eq:alpha1}), yields
\begin{equation}\label{eq:alpha2}
\alpha(b) = - c_1 \log \left( \frac{b}{b_c} - 1 \right) + c_2 + O\left( (b-b_c)\log(b-b_c) \right),
\end{equation}
with $b_c = b(x_m)$ the critical impact parameter: rays with $b > b_c$ escape to infinity, while those with $b < b_c$ spiral into the black hole. From Eqs. (\ref{eq:mfx}) and (\ref{eq:bx0}) we find that 
\begin{equation}
b_c = \sqrt{\frac{C_m}{A_m}} = \frac{\sqrt{2}x_m^2}{\sqrt{x_m + 2\tilde{\gamma}}}.
\end{equation}
Let us apply the general procedure \cite{tsukamoto17}, valid for any spherically symmetric and asymptotically flat geometry with a photon sphere, for obtaining the strong deflection limit coefficients $c_1$ and $c_2$ appearing in Eq. (\ref{eq:alpha2}). The first step in calculating the integral $I(x_0) $ that determines the deflection angle $\alpha$ in Eq. (\ref{eq:alphaint}) is to introduce the new integration variable $z = 1- x_0/x$ and then separate $I(x_0)$ as the sum of two parts $I(x_0) = I_D(x_0) + I_R(x_0)$, where $I_D(x_0)$ diverges logarithmically and $I_R(x_0)$ is regular when $x_0 \to x_m$. The first term $I_D(x_0)$ can be calculated exactly, and the second one $I_R(x_0)$ is only needed at $x_0 = x_m$. By Taylor expanding Eq. (\ref{eq:bx0}) and inverting to obtain $x_0(b)$, both integrals can be written as functions of the impact parameter $b$, to give $I(b)=I_D(b)+I_R(b)$; the first term contains the logarithmic divergence while the second one is regular when  $b \to b_c$. Finally, after some calculations  (see \cite{tsukamoto17} for the details), the strong deflection limit coefficients take the form
\begin{equation}
c_1 = \sqrt{\frac{2 A_m B_m}{C''_m A_m - C_m A''_m}}
\end{equation}
and
\begin{equation}
c_2 = c_1 \log \left[ x_m^2 \left( \frac{C''_m}{C_m} - \frac{A''_m}{A_m} \right) \right] + I_R(x_m) - \pi,
\end{equation}
where the regular integral $I_R(x_m)$ is given by
\begin{equation}\label{eq:ir}
I_R(x_m) = 2 x_m \int_0^1 \ \mathrm{d}z \left[ \sqrt{\frac{B(x(z))}{C(x(z)) R(x(z), x_m)}} \frac{1}{(1-z)^2} - \frac{c_1}{x_m} \frac{1}{z} \right],
\end{equation}
with $R$ defined in Eq. (\ref{eq:r}) and $x(z) = x_m / (1-z)$. The value of $c_1$ is always obtained analytically, while $c_2$ depends on if the integral (\ref{eq:ir}) can be calculated analytically or numerically. In our case, by using Eq. (\ref{eq:mfx}) it is straightforward to see that
\begin{equation} \label{eq:am1}
A_m = \frac{x_m + 2\tilde{\gamma}}{2 x_m^2}, \quad  B_m = \frac{1}{A_m}, \quad C_m = x_m^2 
\end{equation}
and
\begin{equation} \label{eq:am2}
A''_m  = -8 \frac{x_m + 3\tilde{\gamma}}{(3x_m + 4\tilde{\gamma})^2}, \quad C''_m  = 2,
\end{equation}
so, by plugging these equations into the formula for $c_1$, we get
\begin{equation}
c_1 = \frac{\sqrt{2}}{\sqrt{3x_m + 8\tilde{\gamma}}},
\end{equation}
with $x_m$ given by Eq. (\ref{eq:xm}). Turning now to $c_2$, we can see that
\begin{equation}
\left(\frac{B}{CR}\right)^{-1} = \frac{x_m + 2\tilde{\gamma}}{2} \left(\frac{x}{x_m}\right)^4 - x^2 + x + \tilde{\gamma},
\end{equation}
or, in terms of the variable $z$
\begin{equation}
\left(\frac{B}{CR}\right)^{-1} = \frac{x_m + 2\tilde{\gamma}}{2} \frac{1}{(1-z)^4} - \frac{x_m^2}{(1-z)^2} + \frac{x_m}{1-z} + \tilde{\gamma};
\end{equation}
by substituting this equation into the integral (\ref{eq:ir}) and simplifying we get
\begin{equation}
I_R(x_m) = 2 \sqrt{2} x_m \int_0^1 \mathrm{d}z \left[ \frac{1}{z \sqrt{ 2\tilde{\gamma} z^2 - 2(x_m + 4\tilde{\gamma}) z + 3x_m + 8\tilde{\gamma}}} - \frac{1}{\sqrt{3x_m + 8\tilde{\gamma}} z} \right].
\end{equation}
This expression can easily be evaluated using
\begin{equation}
\int \frac{\mathrm{d}z}{z\sqrt{\lambda_2 z^2 + \lambda_1 z + \lambda_0}} = 
\frac{1}{\sqrt{\lambda_0}} \left(\log z - 
\log \left(2\lambda_0 + \lambda_1 z + 2\sqrt{\lambda_0}\sqrt{\lambda_2 z^2 + \lambda_1 z + \lambda_0}\right) \right),
\end{equation}
so, after simplifying, the result is

\begin{figure}[t!]
\centering
\includegraphics[width=0.42\textwidth]{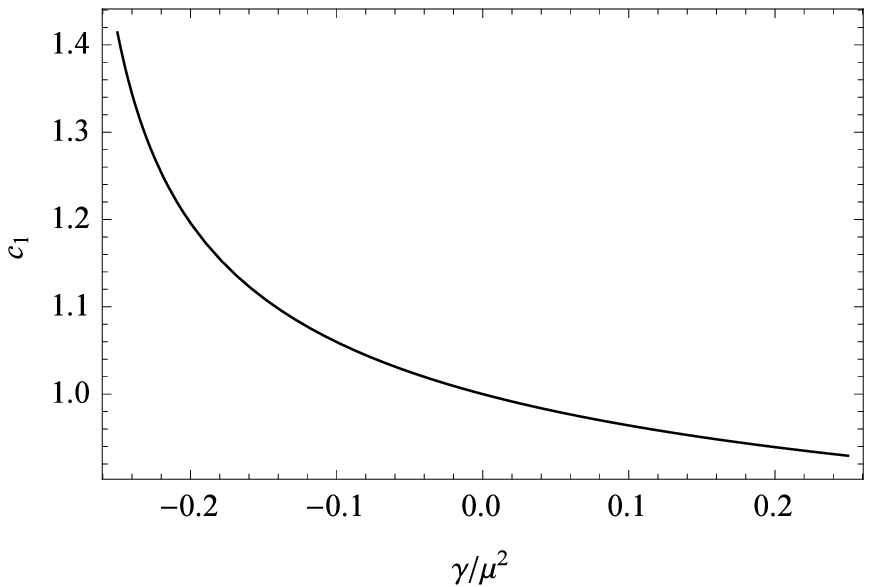}
\hspace{0.05\textwidth}
\includegraphics[width=0.44\textwidth]{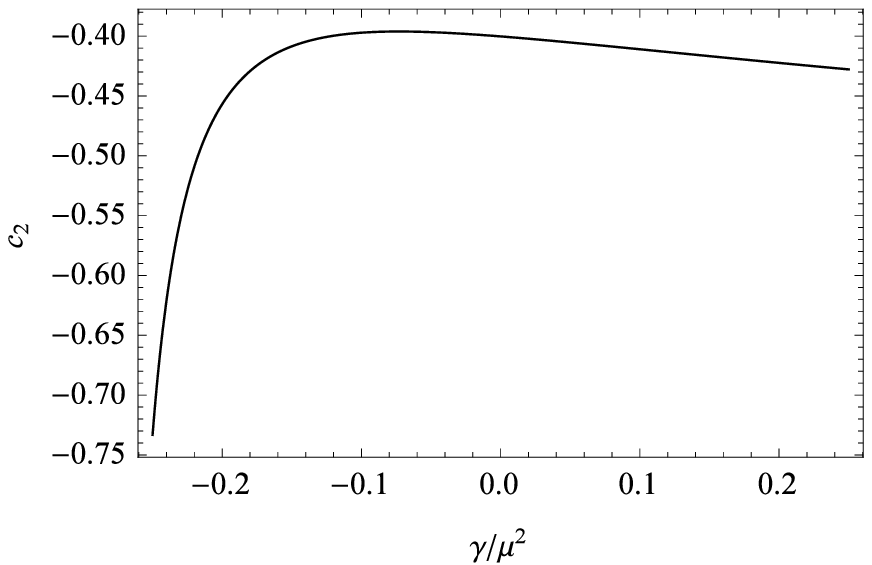}
\caption{The strong deflection limit coefficients $c_1$ (left) and $c_2$ (right) as functions of $\gamma/\mu^2$.}
\label{fig:coef}
\end{figure}

\begin{equation}
I_R(x_m) = c_1 \log
\left[\frac{4(3x_m + 8\tilde{\gamma})^2}{x_m^2(x_m + 2\tilde{\gamma})}
\left(2\sqrt{x_m + 2\tilde{\gamma}} - \sqrt{3x_m + 8\tilde{\gamma}}\right)^2
\right],
\end{equation}
and since
\begin{equation}
\frac{C''_m}{C_m} - \frac{A''_m}{A_m} = 
\frac{2(3x_m + 8\tilde{\gamma})}{x_m^2(x_m + 2\tilde{\gamma})},
\end{equation}
we finally obtain
\begin{equation}
c_2 = c_1 \log
\left[\frac{8(3x_m + 8\tilde{\gamma})^3}{x_m^2(x_m + 2\tilde{\gamma})^2}
\left(2\sqrt{x_m + 2\tilde{\gamma}} - \sqrt{3x_m + 8\tilde{\gamma}}\right)^2
\right] - \pi.
\end{equation}

The values for the Schwarzschild spacetime are recovered with the replacement $\gamma= 0$, i.e. $c_1=1$ and $c_2=\log[216(7-4\sqrt{3})]$. The analytical expressions of the strong deflection limit coefficients $c_1$ and $c_2$ are exact but rather complicated. For a better understanding, their values are plotted  in Fig. \ref{fig:coef} as functions of $\tilde{\gamma} = \gamma/\mu^2$.  We can see that the differences with the corresponding values of the Schwarzschild spacetime are larger for negative $\gamma$ than in the case of positive $\gamma$.

\subsection{Observables}\label{obs}

We consider a point source of light located far behind the black hole and we adopt the observables defined in Ref. \cite{bozza02}. We are interested in the case where there is high alignment between the source and the optical axis (defined as the line connecting the observer and the lens), since the magnifications are largest in this situation \cite{bozza01}. Therefore, we start by writing the deflection angle as $\alpha = 2 \pi n + \Delta \alpha$, with $|\Delta \alpha | \ll 1$. High alignment also implies that the angular positions with respect to the optical axis of the source and the image, $\beta$ and $\theta$ respectively, will be small; the appropriate lens equation for this scenario is \cite{bozza01}
\begin{equation}\label{eq:lens}
\beta = \theta - \frac{D_{LS}}{D_{OS}} \Delta \alpha,
\end{equation}
where $D_{LS}$ is the distance from the lens to the source plane, and $D_{OS}$ the distance from the observer to the source plane. To obtain the image positions, we start by taking the deflection angle in the strong deflection limit approximation and by using the approximate geometric relation $b = \theta D_{OL}$, with $D_{OL}$ the observer-lens distance. Introducing $\alpha = 2 \pi n$ into Eq. (\ref{eq:alpha2}) and solving for $\theta$, we find the solutions
\begin{equation}
\theta_n^0 = \frac{b_c}{D_{OL}}(1 + e_n),
\end{equation}
where
\begin{equation}
e_n = e^{(c_2 - 2n\pi)/c_1}.
\end{equation}
The lens equation (\ref{eq:lens}) can be solved by writing the position of the $n$th image as $\theta_n = \theta_n^0 + \Delta \theta_n$, expanding $\Delta \alpha$ to first order in $\Delta \theta_n$ using Eq. (\ref{eq:alpha2}), and solving for $\Delta \theta_n$; the result is
\begin{equation}
\theta_n = \theta_n^0 + e_n \frac{b_c}{D_{OL}} \frac{D_{OS}}{D_{LS}} \frac{\beta - \theta_n^0}{c_1}
\end{equation}
In this expression we have $e_n \ll 1$ and $b_c/D_{OL} \ll 1$. Although $D_{OS}/D_{LS} \gg 1$, for our purposes it will not be large enough to compensate for the smallness of the other two factors. Therefore, the second term above is much smaller than the first, and it is a good approximation to take $\theta_n \approx \theta_n^0$ when necessary.

Next we calculate the magnifications
\begin{equation}
\mu_n = \left( \frac{\beta}{\theta} \frac{\partial \beta}{\partial \theta} \bigg|_{\theta_n^0} \right) ^{-1},
\end{equation}
which gives
\begin{equation}
\mu_n = e_n \frac{b_c^2 (1 + e_n) D_{OS}}{c_1 \beta D_{OL}^2 D_{LS}}.
\end{equation}
Since the first image is the brightest and has the largest distance from the lens, we assume that it can be distinguished from the rest, which are taken to be packed together at $\theta_\infty = b_c/D_{OL}$, the limiting value of the image positions $\theta_n$. We then define the two observables
\begin{equation}
s = \theta_1 - \theta_\infty \approx \theta_1^0 - \theta_\infty
\end{equation}
and
\begin{equation}
r = \frac{\mu_1}{\sum_{n=2}^\infty \mu_n}
\end{equation}
which after some approximations become
\begin{equation}
s = \theta_\infty e^{(c_2 - 2\pi)/c_1}
\end{equation}
and
\begin{equation}
r = e^{2\pi / c_1}.
\end{equation}

\begin{figure}[t!]
\centering
\includegraphics[width=0.42\textwidth]{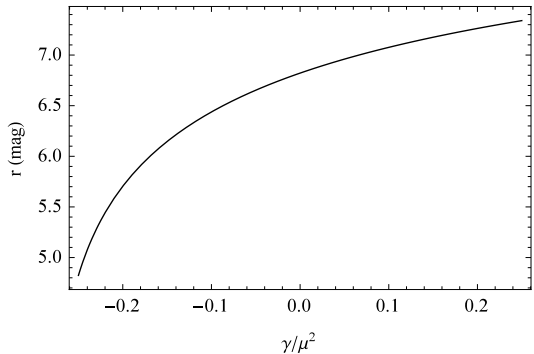}
\caption{Difference in magnitude $r$ between the first relativistic image and the sum of the rest as a function of $\gamma/\mu^2$.}
\label{fig:r}
\end{figure}

\begin{table}[ht]
\begin{center}
\begin{tabular}{|l|c|c|c|c|c|c|c|}
\hline
$\gamma/\mu^2$ & -0.25 & -0.2 & -0.1 & 0 & 0.1 & 0.2 & 0.25 \\[2pt] \hline
$b_c/\mu$ & 2 & 2.16685 & 2.40868 & 2.59808 & 2.75961 & 2.90299 & 2.96956 \\
$c_1$ & 1.41421 & 1.19593 & 1.05964 & 1 & 0.964079 & 0.939274 & 0.92941 \\
$c_2$ & -0.733203 & -0.456856 & -0.397139 & -0.40023 & -0.410979 & -0.422367 & -0.427803 \\ 
$r$ (mag) & 4.8238 & 5.70422 & 6.43791 & 6.82188 & 7.07606 & 7.26293 & 7.34001 \\
\hline
\end{tabular}
\end{center}
\caption{Numerical values of the quantities that are independent of the distance to the black hole: the critical impact parameter $b_c/\mu$, the coefficients $c_1$ and $c_2$, and the observable $r$ (converted to magnitudes), for some representative values of $\gamma/\mu^2$.}
\label{tab:obs}
\end{table}

The observable $s$ is the separation between the first image and the limiting position of the rest, while $r$ is the ratio between the magnification of the first image and the sum of all the others. Since the images come from the same source, $r$ is also the ratio of the received fluxes. We have plotted $r$ as a function of $\gamma/\mu^2$ in Fig. \ref{fig:r}; $s$ and $\theta_\infty$ will be calculated below since they depend on the distance to the chosen black hole. The critical impact parameter $b_c/ \mu $, the coefficients $c_1$ and $c_2$, and $r$ are also summarized in Table \ref{tab:obs}.

\section{Astrophysical applications}\label{astro}

We specialize the general calculations of the previous section to the cases of two supermassive black holes of astrophysical interest: the one in the center of our galaxy and the other in the center of the nearby galaxy M87.

\subsection{Galactic supermassive black hole}

\begin{figure}[t!]
\centering
\includegraphics[width=0.42\textwidth]{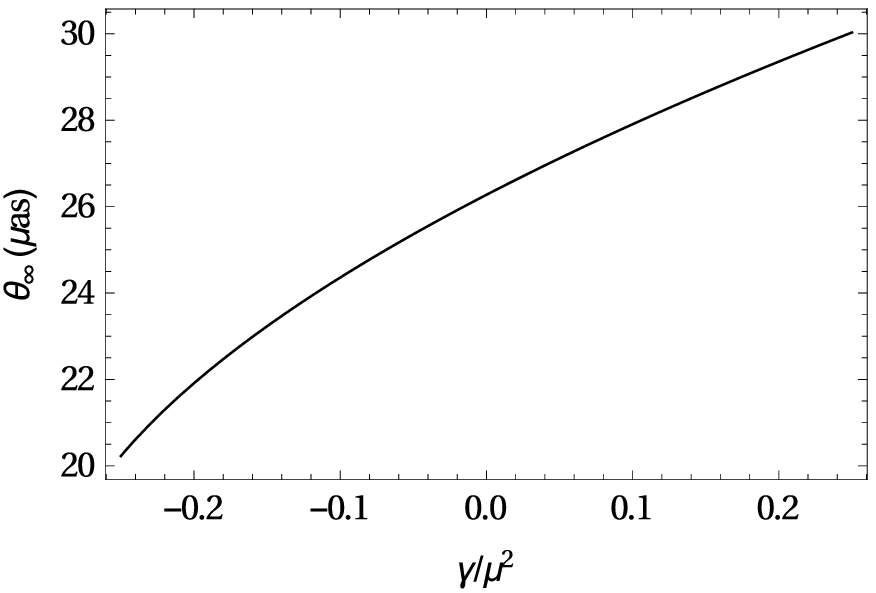}
\hspace{0.05\textwidth}
\includegraphics[width=0.43\textwidth]{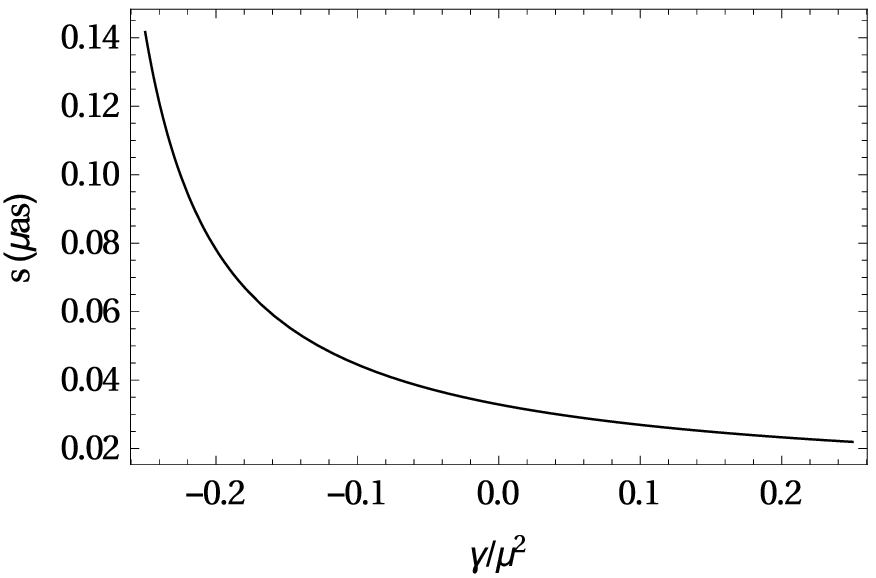}
\caption{Limiting location of the images $\theta_\infty$  (left) and angular separation $s$ between the first image and the limiting value (right), as functions of $\gamma/\mu^2$ for the supermassive black hole in the Milky Way.}
\label{fig:thetaG}
\end{figure}

Firstly, we analyze the case of Sagittarius A*, the supermassive black hole at the center of our galaxy. It seems to be the most promising candidate for the observation of the strong deflection lensing effects. We take the values given in Ref. \cite{gillessen17} of $M = 4.28 \times 10^6 M_{\astrosun}$ and $D_{OL}= 8.32 \, \mathrm{kpc}$. We then have $\mu = 11.9 \times 10^6\, \mathrm{km}$; we must remember that we have chosen $\mu$ as our unit of distance, so that $\tilde{\gamma} = \gamma / \mu^2$, where $\gamma$ has units of length squared. We will also need $D_{OL}/\mu = 2.04 \times 10^{10}$. As explained above, the quantities that depend on the distance to the black hole are $s$ and $\theta_\infty$; they are shown in Fig. \ref{fig:thetaG} and Table \ref{tab:thetaG}.

\begin{table}[ht]
\begin{center}
\begin{tabular}{|l|c|c|c|c|c|c|c|}
\hline
$\gamma/\mu^2$ & -0.25 & -0.2 & -0.1 & 0. & 0.1 & 0.2 & 0.25 \\[2pt] \hline
$\theta _\infty$ ($\mu$as) & 20.2255 & 21.9128 & 24.3583 & 26.2737 & 27.9072 &
29.3572 & 30.0304 \\
$s$ ($\mu$as) & 0.141651 & 0.0781815 & 0.0445364 & 0.0328814 & 0.026925 & 0.0232944 &
0.0219611 \\
\hline
\end{tabular}
\end{center}
\caption{Numerical values of the observables $\theta_\infty$ and $s$ for various values of $\gamma/\mu^2$ for the supermassive black hole in the Milky Way.}
\label{tab:thetaG}
\end{table}

\subsection{Supermassive black hole in M87}

\begin{figure}[t!]
\centering
\includegraphics[width=0.42\textwidth]{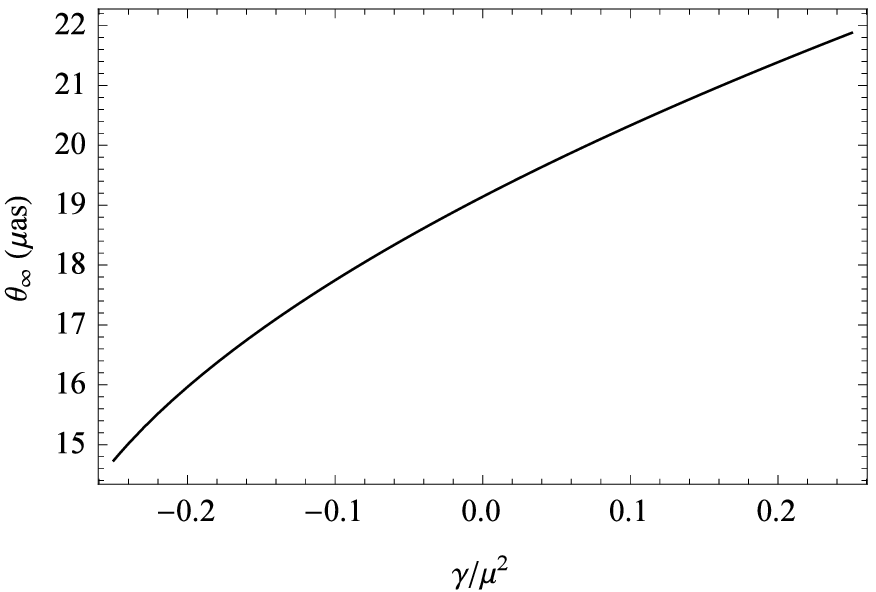}
\hspace{0.05\textwidth}
\includegraphics[width=0.43\textwidth]{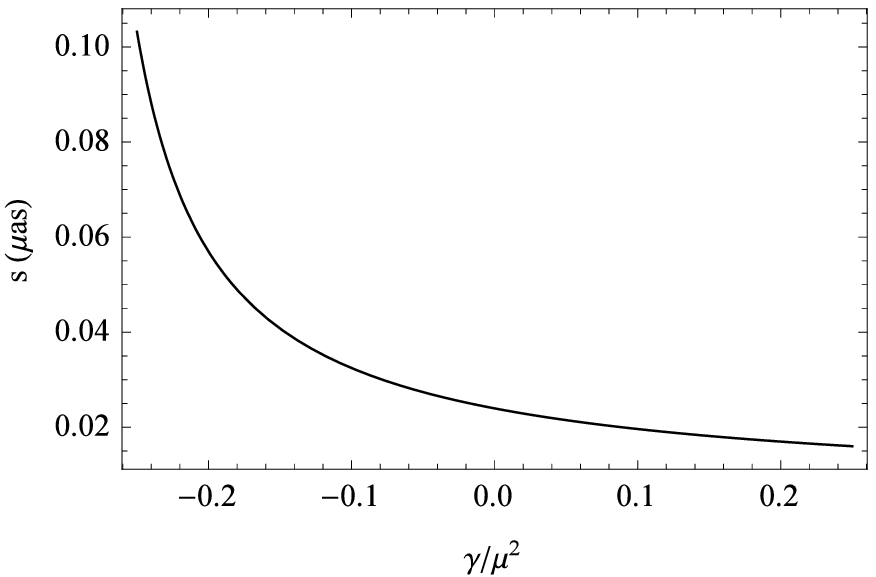}
\caption{Limiting location of the images $\theta_\infty$ (left) and angular separation  $s$ between the first image and the limiting value (right), as functions of $\gamma/\mu^2$ for the supermassive black hole in  M87.}
\label{fig:thetaM87}
\end{figure}

The black hole in the center of the galaxy M87 is also a good candidate for lensing observations; since its mass and distance from Earth are both roughly three orders of magnitude larger than those of Sagittarius A*, the observables are of the same order of magnitude. According to recent studies \cite{broderick15}, this black hole has a mass of $6.16 \times 10^9 M_{\astrosun}$ and is situated at $D_{OL} = 16.5\, \mathrm{Mpc}$ from Earth. Therefore we have $\mu = 1.82 \times 10^{10}\, \mathrm{km}$ and, most importantly, $D_{OL}/\mu = 2.80 \times 10^{10}$, very close to the value for the galactic black hole given above. We display the values of the observables $s$ and $\theta_\infty$ for this black hole in Fig. \ref{fig:thetaM87} and in Table \ref{tab:M87}.

Some care should be taken when comparing Tables \ref{tab:thetaG} and \ref{tab:M87} as well as Figs. \ref{fig:thetaG} and \ref{fig:thetaM87}: since $\gamma$ is a parameter of the theory, $\gamma/\mu^2$ takes different values for the two black holes. In particular, since the mass of the supermassive black hole in M87 is about three orders of magnitude larger than the mass of the one in our galaxy, the corresponding values of $\gamma/\mu^2$ will be about six orders of magnitude smaller.

\begin{table}[ht]
\centering
\begin{tabular}{|l|c|c|c|c|c|c|c|}
\hline
$\gamma/\mu^2$ & -0.25 & -0.2 & -0.1 & 0. & 0.1 & 0.2 & 0.25 \\[2pt] \hline
$\theta _\infty$ ($\mu$as) & 14.7357 & 15.965 & 17.7468 & 19.1423 & 20.3324 & 21.3888 & 21.8793 \\
$s$ ($\mu$as) & 0.103203 & 0.0569608 & 0.0324479 & 0.0239565 & 0.0196168 & 0.0169716 & 0.0160002 \\
\hline
\end{tabular}
\caption{Numerical values of the observables $\theta_\infty$ and $s$ for various values of $\gamma/\mu^2$ for the supermassive black hole in M87.}
\label{tab:M87}
\end{table}

\section{Summary and discussion}\label{sum}

In this work, we have studied gravitational lensing by a black hole in Horndeski theory. This solution, besides the mass $M=\mu/ 2$, depends also on three parameters $\beta$, $\zeta$, and $\eta$. In particular, the metric can be written in terms of $\mu$ and one parameter of the theory $\gamma = \beta ^2/ (2\zeta \eta)$, and it turns out to be identical in form to the Reissner-Nordstr\"om geometry when $\gamma < 0$, by identifying $Q^2= -\gamma$. This parameter $\gamma$ of the theory can, in principle, have any sign, so we have analyzed both possibilities. To obtain an analytical expression for the deflection angle, we have calculated the strong deflection limit coefficients $c_1$ and $c_2$, and, by using the lens equation, the positions and the  magnifications of the relativistic images. We have applied our results to the supermassive black hole at the center of our galaxy and also to the one at the center of the nearby galaxy M87. We have calculated the observables $\theta_\infty$, $s$ and $r$ as functions of $\gamma/ \mu ^2$. These observables are related to the positions and brightnesses of the infinite number of images found at each side of the lens; we assume a hypothetical future observation that can distinguish the first image from the rest. We found that the lensing effects by a Horndeski black hole are stronger than in the Schwarzschild case for $\gamma < 0$, and the differences are smaller for $\gamma > 0$. For $\gamma <0$, a black hole in Horndeski theory can eventually be distinguished from the Reissner-Nordstr\"om geometry by studying the behavior of charged particles in the vicinity of the black hole. An observation with sufficient angular resolution to resolve the first image from all the others would allow one to determine the coefficients $c_1$ and $c_2$. In this way, gravitational lensing can provide an experimental test of General Relativity and alternatives to it, in the strong field region. The precise values of the observables depend on $\gamma/ \mu ^2$, but overall for the Galactic supermassive black hole we expect an angular position of the images of the order of $26 \, \mathrm{\mu as}$, a separation between images of the order of $0.033 \, \mathrm{\mu as}$, and a difference in magnitude of about $6.8\, \mathrm{mag}$. The corresponding values for M87 are similar but smaller. As the parameter $\gamma/ \mu ^2$ determines the lensing effects for the Hondeski black hole, it is of interest to have an idea of which its possible values can be. The parametrized post-Newtonian (PPN) expansion for a metric with the form of Eq. (\ref{eq:metric}) reads \cite{weinberg}  
\begin{equation*}
A(r) = 1-2\alpha_0\frac{M}{r}+2(\beta_0-\alpha_0\gamma_0) \frac{M^2}{r^2}+ ... , 
\end{equation*}
\begin{equation*}
B(r) = 1+2\gamma_0\frac{M}{r}+ ... , \qquad C(r)=r^2.
\end{equation*}
By comparing these expressions with the metric functions given by Eq. (\ref{eq:mfr}), we can easily see that $\alpha_0=\gamma_0=1$ and $\beta_0-\alpha_0\gamma_0=\beta_0-1= -2\gamma/\mu^2$. Currently known solar system limits \cite{will} give $|\beta -1| < 8 \times 10^{-5}$ (perihelion shift of Mercury), so we obtain that $|\gamma|/ \mu ^2 < 4 \times 10^{-5}$. It is not clear if this bound can be extrapolated to the Galactic supermassive black hole and the supermassive black hole in M87. In any case, it seems that if the model adopted here represents a valid description of black holes in nature, the lensing effects will be difficult to distinguish from those corresponding to General Relativity.

We have taken in our model that there is vacuum in vicinity of the black hole. In the case that the black hole is surrounded by a plasma, photons undergo various effects, such as absorption, scattering, and refraction, that depend on the specific characteristics of the medium involved. In the presence of plasma, a shift of angular position of the images, a change in the magnifications, and also chromatic effects can occur \cite{tsupko}.

There are good prospects for the observation of the region close to the event horizon of nearby supermassive black holes in the near future. The instrument GRAVITY will examine in the near-infrared band the vicinity of Sgr A*, following with high precision the orbits of the stars close to Sgr A*. Through this monitoring, the measurement of the black hole mass will be improved, and the determination of the spin and the quadrupole moment may also be possible. The Event Horizon Telescope, a very long baseline interferometry network of millimeter and sub-millimeter wavelength instruments, is expected to obtain the first direct image of nearby supermassive black holes. This is of great importance since the size and shape of the shadow depends directly on the properties of the corresponding spacetime, allowing to test General Relativity and alternatives to it. Forthcoming  x-ray instruments  are also expected to have a better resolution, in order to achieve a more detailed exploration of the Galactic center in this band.  A comprehensive treatment of this topic can be found in the literature (e.g. \cite{observ} and the references therein). However, the observation of the subtle effects like those discussed here seems to be outside the capabilities of these facilities and will require a more advanced generation of instruments.

\end{document}